\documentclass[amsmath,amssymb,
onecolumn,apm]{revtex4-2}

\usepackage{graphicx}
\usepackage{caption}
\usepackage{xcolor}
\usepackage[utf8]{inputenc}
\usepackage{rotating}

\DeclareUnicodeCharacter{2212}{-}

\begin{document}

\title{Dynamical interplay between coupled scalar dark sectors and gravity}

\author{Mihai Marciu}
\email{mihai.marciu@drd.unibuc.ro}
\affiliation{ 
Faculty of Physics, University of Bucharest, Bucharest-Magurele, Romania
}

\date{\today}

\begin{abstract}
We explore a novel cosmological model based on coupled fields in the framework of scalar tensor theories, considering the specific interplay between gravity and scalar fields. The model further extends a recent axion-dilaton system by introducing viable couplings with the space--time geometry encoded into the scalar curvature. After briefly introducing the action and the corresponding field equations, we employ linear stability theory to investigate the physical properties. The analysis showed the compatibility of the current theoretical model with the recent history of the Universe, obtaining viable constraints for the model's parameters in some specific cases. In the present setup, the axion--dilaton system is non--minimally coupled with gravity in an independent manner, leading to distinct physical features in the phase-space structure, possible alleviating the cosmic coincidence problem. 
\end{abstract}

\maketitle

\section{Introduction}
\label{intro}

\par 
In the modern cosmological theories one of the greatest questions is related to the origin of the accelerated expansion of the Universe \cite{Copeland:2006wr, Peebles:2002gy}. This question has driven the opening of various theoretical directions \cite{Nojiri:2010wj, SupernovaSearchTeam:2004lze}, having ramifications in fundamental science. In these theories, the dark sector is composed by the dark energy and dark matter \cite{Planck:2018vyg}, dominating the cosmic picture, affecting the evolution of the baryonic matter \cite{Padmanabhan:2002ji}. 
\par 
The dark matter \cite{Navarro:1995iw, Arkani-Hamed:2008hhe} represents another mysterious component of the Universe, affecting the dynamics of the Universe on local scales \cite{Bertone:2004pz, Clowe:2006eq, LUX:2016ggv}. This component is having an equation of state corresponding to a cold or almost cold fluid, surrounding the baryonic matter on galactic scales \cite{Bertone:2004pz, Clowe:2006eq}. The interdependence and tracking between ordinary matter and dark matter in the history of our Universe opens various directions for future studies \cite{Davis:1985rj}. The dark matter has been probed and analyzed by various astrophysical studies \cite{Planck:2015fie, WMAP:2003elm, Navarro:1996gj, WMAP:2010qai, BOSS:2016wmc}. Beyond galactic scales, at large scale structure, the dynamics is governed by the dark energy, another unknown component which drives the acceleration of the Universe, affecting the cosmic boundaries \cite{Bamba:2012cp}. This effect has been studied intensively in the past years since the discovery at the end of the last millennium \cite{Perivolaropoulos:2021jda}. From a technological point of view, can be regarded as a limitless source of energy, a mysterious physical phenomenon connected to the evolution of the Universe \cite{Amendola:2016saw}.
\par 
The most simple model is the $\Lambda$CDM model \cite{Perivolaropoulos:2021jda}, where the Einstein's equations are encapsulated with a cosmological constant term, acting as an effective theory. Although this theory can in principle explain the accelerated expansion, different evolutionary aspects of the dark energy - dark matter system are not elucidated \cite{Perivolaropoulos:2021jda}. The main problem of the $\Lambda$CDM model \cite{Bull:2015stt} is observed when treating the dark energy component as a field - the specific equation of state evolves dynamically with respect to the redshift or cosmic time, leading to a discrepancy between theoretical aspects and astrophysical observations \cite{SupernovaSearchTeam:2003cyd, SupernovaCosmologyProject:2011ycw, eBOSS:2020yzd, Amanullah:2010vv, DES:2017qwj}. Moreover, a phantom evolution \cite{Carroll:2003st, Nojiri:2005sx, Andriot:2025los, Fikri:2024klc} seems to be favored by observations, an intriguing result overthrowing various classical aspects of theories \cite{Ludwick:2017tox}.      
\par 
Another compatible attempt is represented by modified gravity theories \cite{Clifton:2011jh, Nojiri:2017ncd, Joyce:2014kja}, where different Lagrangians have been proposed, extending the Einstein-Hilbert action in a non-trivial manner. These theories are based on various invariants as geometrical or non-geometrical components \cite{Koyama:2015vza}. On one hand the geometrical invariants can be associated to the scalar curvature \cite{Sotiriou:2008rp, DeFelice:2010aj}, the Gauss-Bonnet term \cite{Nojiri:2005jg, Cognola:2006eg, Marciu:2019cpb}, or other invariants in higher order theories \cite{Ali:2025ybt, Marciu:2020ysf, Marciu:2020ski, Marciu:2022wzh, Marciu:2021rdl, Marciu:2023hdb, Marciu:2022rsc}. On the other hand, the non-geometrical invariants can include influences due to various properties of the matter component, based on specific contractions associated to the energy-momentum tensor \cite{Harko:2011kv, Bahamonde:2019urw, Cipriano:2024jng}. In the recent years different ideas have been explored, taking into account that matter and geometry can be on more or less equal footing \cite{Marciu:2024gqv, Bouali:2023fid, Pinto:2022tlu}, creating or annihilating matter due to the interplay between matter and geometry \cite{Cipriano:2023yhv}.
\par 
In the framework of modified gravity theories a special class is dedicated to the scalar fields \cite{Ishak:2018his}, where the accelerated expansion is associated to the evolution of a scalar component which can be minimally or non-minimally coupled to various invariants having a geometric origin \cite{Bahamonde:2017ize}. The scalar fields are classified also by specific properties due to the form of the kinetic energy and potential. The kinetic energy describes the phantom and quintessence behavior in the single or multi scalar field models \cite{Ishak:2018his}. In multi scalar cosmologies \cite{Heisenberg:2018vsk} one can have a superposition of scalar fields, with non negligible effects from the kinetic and potential energy. A special class named quintom \cite{Guo:2004fq,Marciu:2016aqq, Marciu:2018oks} is defined by the superposition between a phantom field and a quintessence one, leading to the crossing of the phantom divide line in the recent past, a viable alternative in accordance to astrophysical observations. 
\par 
The coupling between the scalar fields and various geometrical invariants has been considered in the past years \cite{Hrycyna:2020jmw,Hrycyna:2015eta, Hrycyna:2010yv}, indicating a compatibility with physical observations \cite{Hrycyna:2015vvs}. Furthermore, this coupling can lead to various scaling solutions \cite{Hrycyna:2009zj, Marciu:2020vve} which can alleviate the cosmic coincidence problem. 
\par 
In view of these considerations, we have extended the previous dark sector \cite{Rahimy:2025iyj} based on two scalar fields, a saxion and an axion component, by including viable influences from the geometrical manifold. Hence, the paper represents a particular extension of the minimal coupling model \cite{Rahimy:2025iyj} towards a more complete theory of gravity.
\par 
In the present paper we shall explore the effects due to the interplay between the scalar fields and the geometrical manifold, expressed through the non-minimal coupling of the scalar fields with the scalar curvature. The physical effects are analyzed by considering the dynamical analysis. This analytical tool \cite{Bahamonde:2017ize} is important in the study of cosmological systems, exposing the richness or limitations of the phase space structure. The linear stability theory has been applied to various models in the recent times \cite{Bahamonde:2017ize}. 
\par 
The paper is organized as follows. In Sec.~\ref{actiune} we present the action for the current model, obtaining the corresponding field equations. Then, in Sec.~\ref{phase} we analyze the physical implications by applying the linear stability theory. Lastly, in Sec.~\ref{conclusions} we make a short summary and the final concluding remarks.

\section{The action and the field equations}
\label{actiune}

\par 
In what follows we shall consider that the dark sector can be represented by two scalar fields which are non--minimally coupled with gravity through the scalar curvature. Hence, the action that we are proposing have the following form:

\begin{equation}
    \label{act}
    S=\int d^4 x \sqrt{-g} \Big[ \frac{R}{2} -  \frac{\epsilon_1}{2} \partial_{\mu} \phi \partial^{\mu} \phi - \frac{\epsilon_2 f^2 (\phi)}{2} \partial_{\mu} \chi \partial^{\mu} \chi - V(\phi, \chi) + g_1 (\phi) R +g_2 (\chi) R \Big]+ S_m,
\end{equation}
where $g$ represents the determinant of the metric, $R$ the scalar curvature, and $S_m$ the matter action, acting independently. In this case the Einstein--Hilbert action is extended by considering two scalar fields having specific kinetic energies and a generalized potential. We note that the kinetic energy of the $\chi$ field is modulated through a specific function related to the value of the scalar field $\phi$. As in Ref.~\cite{Rahimy:2025iyj}, we shall consider that the $\phi$ field is a saxion--like field, while $\chi$ represents an axion--like component. The mixture between the kinetic energy and the dynamics of the scalar field in the action is investigated by using the dynamical system approach. Before proceeding to the direct analysis, we briefly describe the metric that is considered. We shall work with the usual Robertson--Walker metric expressed as:

\begin{equation}
\label{metr}
    ds^2=-dt^2+a^2(t)dx_i dx^i,
\end{equation}

where $a(t)$ represents the associated scale factor which depends on cosmic time, describing a homogeneous and isotropic cosmological scenario known as the Friedmann–Lemaître–Robertson–Walker model. The variation of the action \eqref{act} with respect to the metric \eqref{metr} gives the following modified Friedmann relations:

\begin{equation}
    3 H^2=\rho_m+\frac{\epsilon_1}{2}\dot{\phi}^2+V(\phi, \chi)+\frac{\epsilon_2}{2}f^2(\phi)\dot{\chi^2}-6 g_1(\phi)H^2-6 H g_1'(\phi)\dot{\phi}-6 g_2(\chi)H^2-6 H g_2'(\chi)\dot{\chi},
\end{equation}

\begin{multline}
    -3H^2-2\dot{H}=\frac{\epsilon_1}{2}\dot{\phi}^2-V(\phi, \chi)+\frac{\epsilon_2}{2}f^2(\phi)\dot{\chi^2}+6 g_1(\phi)H^2+4 g_1(\phi) \dot{H}+4 H g_1'(\phi)\dot{\phi}+2 \dot{\phi}^2 g_1''(\phi)+2 g_1'(\phi)\ddot{\phi}
    \\+6 g_2(\chi)H^2+4 g_2(\chi) \dot{H}+4 H g_2'(\chi)\dot{\chi}+2 \dot{\chi}^2 g_2''(\chi)+2 g_2'(\chi)\ddot{\chi}.
\end{multline}

and the next Klein--Gordon equations:

\begin{equation}
    \epsilon_1 (\ddot{\phi}+3 H \dot{\phi})+\frac{dV}{d \phi}=\epsilon_2 f(\phi) f'(\phi) \dot{\chi}^2+6 g_1'(\phi)(\dot{H}+2 H^2),
\end{equation}

\begin{equation}
    \epsilon_2 (\ddot{\chi}+3 H \dot{\chi})+\frac{1}{f^2(\phi)}\frac{dV}{d \chi}=-2 \epsilon_2 \frac{f'(\phi)}{f(\phi)}\dot{\phi}\dot{\chi}+6 g_2'(\chi)(\dot{H}+2 H^2)\frac{1}{f^2(\phi)}.
\end{equation}

\par 
For this system we can check the viability of the standard continuity relation, 
\begin{equation}
    \dot{\rho_{\phi \chi}}+3 H (\rho_{\phi \chi} +p_{\phi \chi})=0,
\end{equation}

\begin{equation}
    \rho_{\phi \chi}=\frac{\epsilon_1}{2}\dot{\phi}^2+V(\phi, \chi)+\frac{\epsilon_2}{2}f^2(\phi)\dot{\chi^2}-6 g_1(\phi)H^2-6 H g_1'(\phi)\dot{\phi}-6 g_2(\chi)H^2-6 H g_2'(\chi)\dot{\chi},
\end{equation}

\begin{multline}
    p_{\phi \chi}=\frac{\epsilon_1}{2}\dot{\phi}^2-V(\phi, \chi)+\frac{\epsilon_2}{2}f^2(\phi)\dot{\chi^2}+6 g_1(\phi)H^2+4 g_1(\phi) \dot{H}+4 H g_1'(\phi)\dot{\phi}+2 \dot{\phi}^2 g_1''(\phi)+2 g_1'(\phi)\ddot{\phi}
    \\+6 g_2(\chi)H^2+4 g_2(\chi) \dot{H}+4 H g_2'(\chi)\dot{\chi}+2 \dot{\chi}^2 g_2''(\chi)+2 g_2'(\chi)\ddot{\chi},
\end{multline}
since the model doesn't imply any interaction with the energy momentum tensor of the matter component beyond the standard level.
\par 
Finally, we can define the energy density of the dark sector, and the energy density of the matter component as follows, 

\begin{equation}
    \Omega_{\phi \chi}=\frac{\rho_{\phi \chi}}{3 H^2},
\end{equation}

\begin{equation}
    \Omega_m=\frac{\rho_m}{3 H^2},
\end{equation}

\begin{equation}
    \Omega_{\phi \chi}+ \Omega_m=1.
\end{equation}

\section{The phase space analysis}
\label{phase}
\par 
In order to analyze the dynamical consequences for the coupling with gravity through the curvature scalar, we shall consider using the basic linear stability theory. This enables us to discover the complexity of the phase-space structure, obtaining possible constraints from a theoretical point of view. Before introducing the dimension--less variables, we shall describe the coupling terms. For the analysis, we have considered the following exponential coupling terms and decompositions:

\begin{equation}
    V(\phi, \chi)=W(\phi)+\tilde{g}(\phi)U(\chi),
\end{equation}

\begin{equation}
  W(\phi)=W_0 e^{- \lambda_2 \phi},  
\end{equation}

\begin{equation}
  \tilde{g}(\phi)=g_0 e^{- \gamma \phi},  
\end{equation}

\begin{equation}
  U(\chi)=U_0 e^{- \lambda_1 \chi},  
\end{equation}

\begin{equation}
  f(\phi)=f_0 e^{+ \beta \phi},  
\end{equation}

\begin{equation}
  g_1(\phi)=g_{1,0} e^{\alpha_1 \phi},  
\end{equation}

\begin{equation}
  g_2(\chi)=g_{2,0} e^{\alpha_2 \chi}.  
\end{equation}

\par 

Next, for the analysis, we shall introduce the following dimension--less variables \cite{Rahimy:2025iyj, Licciardello:2025fhx}:

\begin{equation}
    x_1=\frac{\dot{\chi}}{\sqrt{6}H},
\end{equation}

\begin{equation}
    x_2=\frac{\dot{\phi}}{\sqrt{6}H},
\end{equation}

\begin{equation}
    x_f=\frac{f(\phi)\dot{\chi}}{\sqrt{6}H},
\end{equation}

\begin{equation}
    y_1=\frac{\tilde{g}(\phi)U(\chi)}{3 H^2},
\end{equation}

\begin{equation}
    y_2=\frac{W(\phi)}{3 H^2},
\end{equation}

\begin{equation}
    u=\frac{\rho_m}{3 H^2}
\end{equation}

\begin{equation}
    z_1=2 g_2(\chi)
\end{equation}

\begin{equation}
    z_2=2 g_1(\phi).
\end{equation}

\par 
Next, we introduce the e-folding number $N=log(a)$, changing the dependence from the cosmic time, obtaining the following system of equations in the first order:

\begin{equation}
    \frac{dx_1}{dN}=-x_1 \frac{\dot{H}}{H^2}+\frac{\ddot{\chi}}{\sqrt{6}H^2},
\end{equation}

\begin{equation}
    \frac{dx_2}{dN}=-x_2 \frac{\dot{H}}{H^2}+\frac{\ddot{\phi}}{\sqrt{6}H^2},
\end{equation}

\begin{equation}
    \frac{d x_f}{dN}=\beta \sqrt{6} x_2 x_f-x_f \frac{\dot{H}}{H^2}+x_f\frac{\ddot{\chi}}{\sqrt{6} x_1 H^2 },
\end{equation}

\begin{equation}
    \frac{dy_2}{dN}=-\sqrt{6} \lambda_2 x_2 y_2-2 y_2\frac{\dot{H}}{H^2},
\end{equation}

\begin{equation}
    \frac{du}{dN}=-3u-2u\frac{\dot{H}}{H^2},
\end{equation}

\begin{equation}
    \frac{dz_1}{dN}=\sqrt{6} \alpha_2 x_1 z_1,
\end{equation}

\begin{equation}
    \frac{dz_2}{dN}=\sqrt{6}\alpha_1 x_2 z_2.
\end{equation}

\par 
Note that we used the Friedmann constraint equation in order to reduce the dimensionality of the autonomous system, replacing the $y_1$ variable with the following expression:

\begin{equation}
    y_1 = -\epsilon _2 x_f^2-u+\sqrt{6} \alpha _1 x_2 z_2+\sqrt{6} \alpha _2 x_1 z_1-x_2^2 \epsilon _1-y_2+z_1+z_2+1.
\end{equation}

\par 
The Klein-Gordon equation for the $\phi$ field reduces to:

\begin{equation}
    \ddot{\phi}= \frac{3 \left(H^2 \left(2 \beta  \epsilon _2 x_f^2-\sqrt{6} x_2 \epsilon _1+\gamma  y_1+\lambda _2 y_2+2 \alpha _1 z_2\right)+\alpha _1 z_2 \dot{H}\right)}{\epsilon _1},
\end{equation}

while for the $\chi$ field we have:

\begin{equation}
    \ddot{\chi}= \frac{3 x_1 \left(H^2 \left(x_1 \left(\lambda _1 y_1+2 \alpha _2 z_1\right)-\epsilon _2 x_f^2 \left(4 \beta  x_2+\sqrt{6}\right)\right)+\alpha _2 x_1 z_1 \dot{H}\right)}{\epsilon _2 x_f^2}.
\end{equation}

\par 
The acceleration equation for our cosmological system can be written as:

\begin{multline}
    6 H^2 \epsilon _2 x_f^2+12 \alpha _1^2 H^2 x_2^2 z_2+12 \alpha _2^2 H^2 x_1^2 z_1+6 H^2 x_2^2 \epsilon _1-2 \left(3 H^2 y_1+3 H^2 y_2\right)+6 H^2 \left(z_1+z_2+1\right)+4 z_1 \dot{H}+4 z_2 \dot{H}+4 \dot{H}
    \\+8 H \left(\sqrt{\frac{3}{2}} \alpha _1 H x_2 z_2+\sqrt{\frac{3}{2}} \alpha _2 H x_1 z_1\right)+2 \alpha _2 z_1 \ddot{\chi}+2 \alpha _1 z_2 \ddot{\phi}=0.
\end{multline}

\par 
For this system we have obtained the corresponding relations for $\Big[\frac{\ddot{\phi}}{H^2}, \frac{\ddot{\chi}}{H^2}, \frac{\dot{H}}{H^2}\Big]$, deducing the final expressions for the autonomous system of equations in the first order. For the two fields we have taken into consideration that the kinetic energies have positive signs, as canonical fields ($\epsilon_{1,2}=+1$). In order to apply the dynamical system analysis, we have equalized the right hand side of the autonomous system to zero, obtaining the corresponding critical points which are specified in Table I. In what follows, we shall discuss each critical point obtained in detail, analyzing the phase space complexity with respect to the cosmological evolution of our Universe.
\par
\begin{figure}[t]
  \includegraphics[width=8cm]{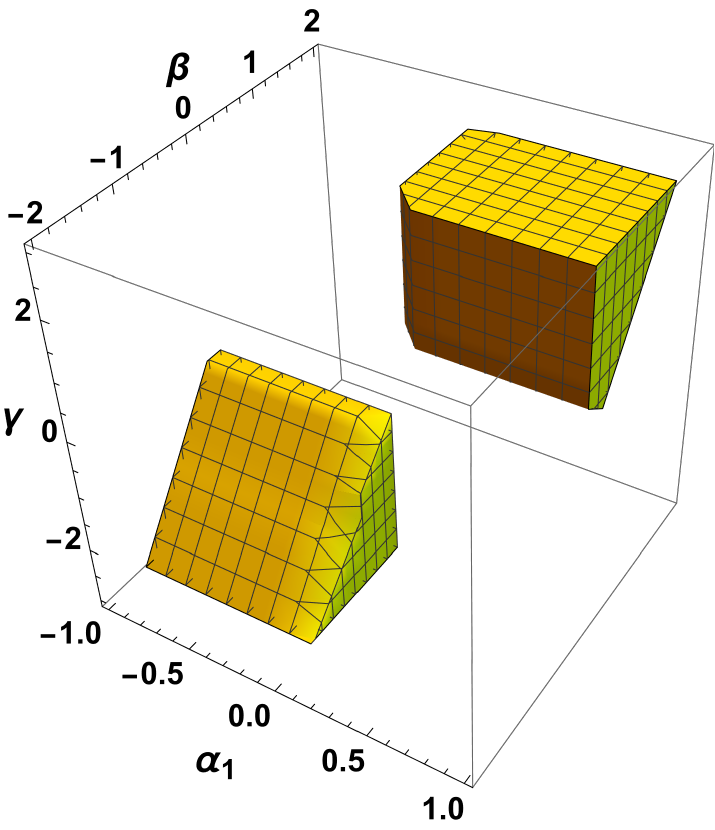} 
\caption{The figure describes a possible saddle region for the $P_1$ critical line.}
\label{fig:p1s}       
\end{figure}

\begin{figure}[t]
  \includegraphics[width=8cm]{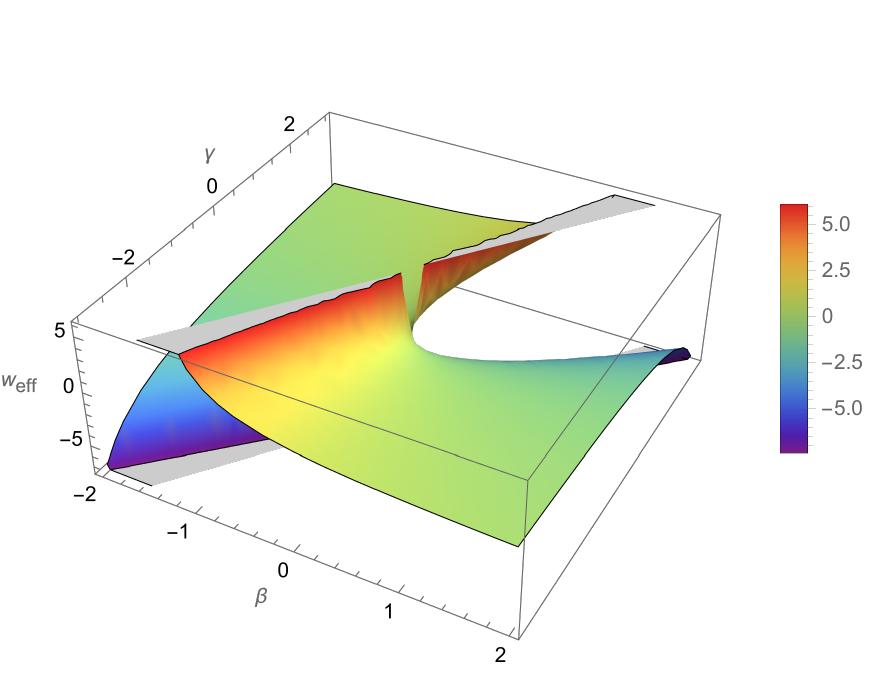} 
\caption{The effective equation of state for the $P_1$ solution.}
\label{fig:p1weff}       
\end{figure}

 
The first solution denoted as $P_1$ is located at the following coordinates:

\begin{equation}
    P_1=\Big[x_1\to 0,x_2\to \frac{\sqrt{6}}{\gamma -2 \beta },y_2\to 0,u\to 0,z_1\to \frac{-2 \beta  \gamma +\gamma ^2-4 \beta ^2 x_f^2+4 \beta  \gamma  x_f^2+\gamma ^2 \left(-x_f^2\right)-6}{\gamma  (2 \beta -\gamma )},z_2\to 0 \Big],
\end{equation}
with the effective equation of state, 
\begin{equation}
    w_{eff}=-\frac{2 \beta +\gamma }{2 \beta -\gamma }.
\end{equation}
The eigenvalues for this solution have the following form:
\begin{multline}
    P_1^{E}=\Bigg[ 0,\frac{6 \alpha _1}{\gamma -2 \beta },-\frac{6 \beta }{\gamma -2 \beta },\frac{12 \beta }{\gamma -2 \beta }+3,-\frac{6 \left(\gamma -\lambda _2\right)}{2 \beta -\gamma },
    \\
    \frac{\pm \sqrt{3} \sqrt{-\beta  \gamma ^2 (\gamma -2 \beta )^2 x_f^2 \left(2 (2 \beta -\gamma )^3 x_f^2-3 \beta \right) \left((\gamma -2 \beta )^2 x_f^2+6\right){}^2}-3 \beta  \gamma  (2 \beta -\gamma )^3 x_f^3+18 \beta  \gamma  (\gamma -2 \beta ) x_f}{\gamma  (\gamma -2 \beta )^2 x_f \left((\gamma -2 \beta )^2 x_f^2+6\right)} \Bigg].
\end{multline}
Since one eigenvalue is equal to zero, the solution is non--hyperbolic and the linear stability theory becomes limited in evaluating the dynamical aspects for this critical point. In Fig.\ref{fig:p1s} we have depicted a region where the $P_1$ solution is saddle, due to the existence of an at least one positive eigenvalue, followed by also an at least one negative eigenvalue. In Fig.\ref{fig:p1weff} we have plotted the effective equation of state for the $P_1$ solution. As can be seen, this solution can describe many cosmological eras, spanning also the phantom region.

\begin{figure}[t]
  \includegraphics[width=8cm]{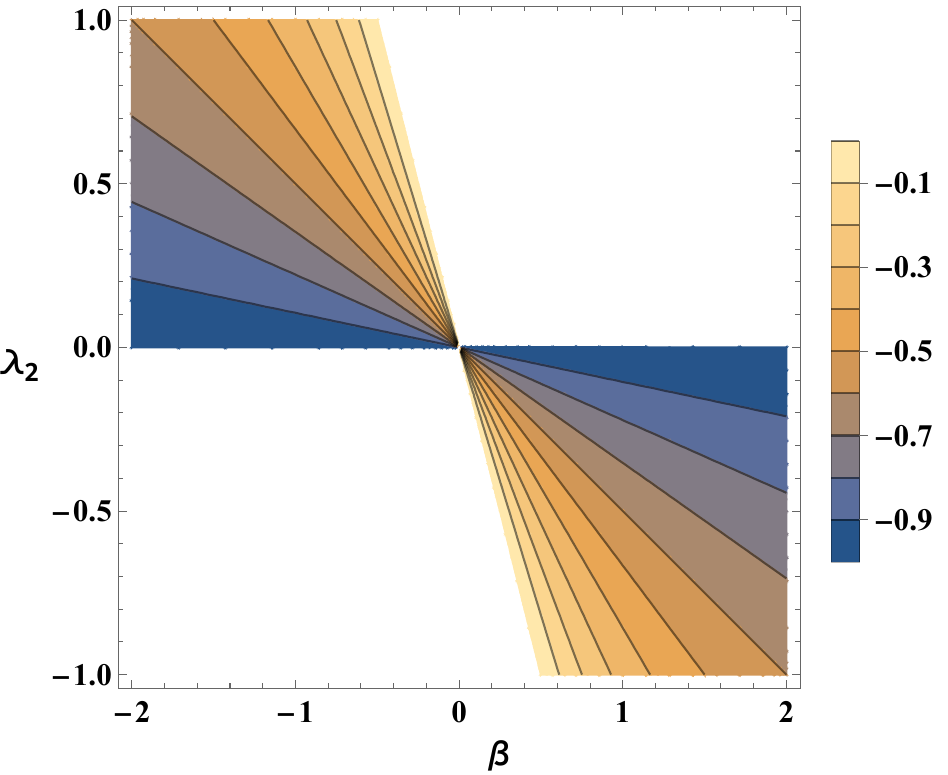} 
\caption{The effective equation of state for the $P_2$ solution.}
\label{fig:p2weff}       
\end{figure}

\par 
The second solution can be found at the coordinates, 

\begin{multline}
    P_2=\Big[x_1\to 0,x_2\to -\frac{\sqrt{6}}{2 \beta -\lambda _2},y_2\to -\frac{2 \left(6 \beta +4 \beta ^3 x_f^2-4 \beta ^2 \lambda _2 x_f^2+\beta  \lambda _2^2 x_f^2\right)}{\lambda _2 \left(2 \beta -\lambda _2\right){}^2},u\to 0,
    \\
    z_1\to \frac{-2 \beta  \lambda _2-4 \beta ^2 x_f^2+4 \beta  \lambda _2 x_f^2-\lambda _2^2 x_f^2+\lambda _2^2-6}{\lambda _2 \left(2 \beta -\lambda _2\right)},z_2\to 0 \Big],
\end{multline}
with the effective equation of state, 
\begin{equation}
    w_{eff}=-\frac{2 \beta +\lambda _2}{2 \beta -\lambda _2},
\end{equation}
describing a cosmological scenario where the geometrical dark energy of the axion-saxion system dominates. As can be noted, the kinetic energy of the $\phi$ field is affecting the dynamics, together with its potential, complimented by the scalar curvature coupling of the $\chi$ field. The eigenvalues for this solution are the following:

\begin{multline}
    P_2^{E}=\Bigg[ 0,\frac{6 \alpha _1}{\lambda _2-2 \beta },-\frac{6 \beta }{\lambda _2-2 \beta },\frac{6 \left(\gamma -\lambda _2\right)}{2 \beta -\lambda _2},\frac{12 \beta }{\lambda _2-2 \beta }+3,
    \\
    \frac{-3 \beta  \lambda _2 \left(\lambda _2-2 \beta \right){}^4 x_f^3-18 \beta  \lambda _2 \left(\lambda _2-2 \beta \right){}^2 x_f\pm\sqrt{3} \sqrt{-\beta  \lambda _2^2 \left(\lambda _2-2 \beta \right){}^4 x_f^2 \left(2 \left(2 \beta -\lambda _2\right){}^3 x_f^2-3 \beta \right) \left(\left(\lambda _2-2 \beta \right){}^2 x_f^2+6\right){}^2}}{\lambda _2 \left(2 \beta -\lambda _2\right){}^3 x_f \left(\left(\lambda _2-2 \beta \right){}^2 x_f^2+6\right)}\Bigg].
\end{multline}
For this solution the effective equation of state can have various values, corresponding to different cosmological eras, as for the $P_1$ solution. This can be observed in Fig.\ref{fig:p2weff}.
\par 
The next solution $P_3$ describes a critical line at the coordinates described by:

\begin{multline}
    P_3=\Big[ x_1\to \frac{\sqrt{\frac{3}{2}}}{\lambda _1},x_2\to 0,y_2\to 0,u\to \frac{\alpha _1-2 \alpha _1 x_f^2-4 \beta  x_f^2-2 \gamma  x_f^2}{\alpha _1},z_1\to 0,z_2\to -\frac{2 (2 \beta +\gamma ) x_f^2}{\alpha _1} \Big],
\end{multline}

with a zero effective equation of state $w_{eff}=0$, corresponding to a scaling solution which in principle can alleviate the cosmological coincidence problem. For this solution the kinetic energy of the axion-like field is non-zero, affecting the location in the phase space structure, together with coupling strength of the saxion-like field. For the eigenvalues, we have obtained the following expressions:
\begin{equation}
    P_3^{E}=\Big[0,\frac{3 \alpha _2}{\lambda _1},3, ... \Big].
\end{equation}
Note that the last eigenvalues in this case have complicated expressions and are not displayed in the manuscript. As can be noted, one eigenvalue is always real and positive, implying that this solution can be only saddle or unstable. For the alleviation of cosmic coincidence problem the scaling solution should be saddle. In the case when we fix $x_f=1, \beta=1, \alpha_1=1, \gamma=1$ the complexity of the eigenvalues is reduced, 

\begin{equation}
    P_3^{E}=\Big[0,\frac{3 \alpha _2}{\lambda _1},3,\pm \frac{\sqrt{12 \sqrt{11471} \sqrt{-\lambda _1^{12}}-939 \lambda _1^6}}{28 \lambda _1^3}-\frac{3}{4},\pm \frac{\sqrt{3} \sqrt{-313 \lambda _1^6-4 \sqrt{11471} \sqrt{-\lambda _1^{12}}}}{28 \lambda _1^3}-\frac{3}{4} \Big].
\end{equation}

\begin{figure}[t]
  \includegraphics[width=8cm]{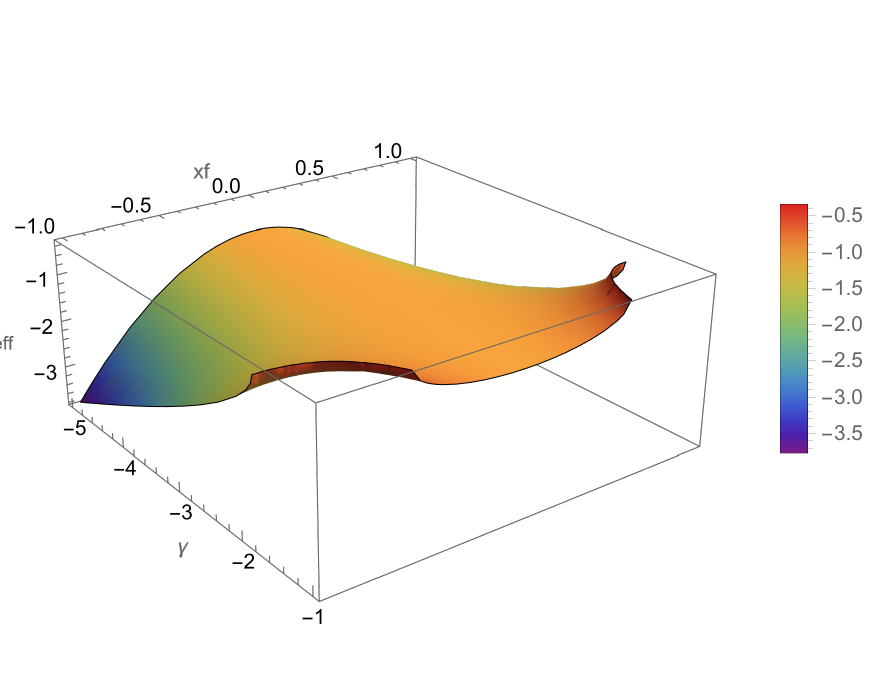} 
\caption{The effective equation of state for the $P_4^{+}$ solution ($\beta=1, \lambda_1=1, \alpha_1=1$).}
\label{fig:p4plusweff}       
\end{figure}

\begin{figure}[t]
  \includegraphics[width=8cm]{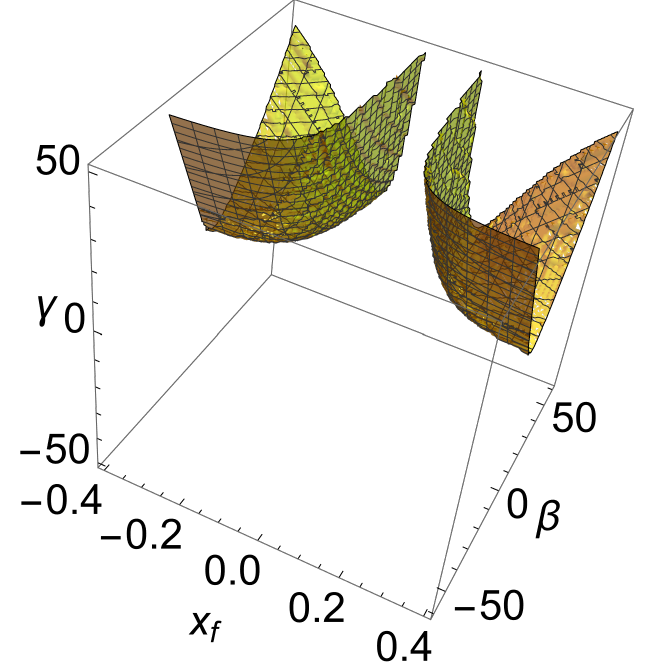} 
\caption{The contour for the $P_4^{+}$ solution where $w_{eff}=0$ ($\alpha_1=1, \lambda_1=1$).}
\label{fig:p4plusweffzero}       
\end{figure}

\begin{figure}[t]
  \includegraphics[width=8cm]{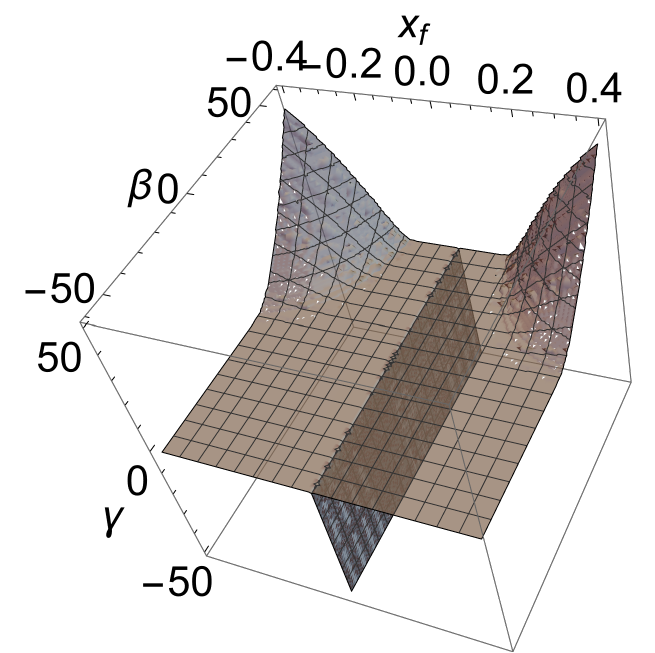} 
\caption{The contour of the effective equation of state for the $P_4^{+}$ solution where $w_{eff}=-1$ ($\lambda_1=1, \alpha_1=1$).}
\label{fig:p4plusweffminusunu}       
\end{figure}

The $P_4$ solution located at the coordinates:

\begin{multline}
    P_4^{\pm}=\Bigg[x_1\to \frac{2 \alpha _1 \lambda _1+\lambda _1 x_f^2 \left(3 \alpha _1-2 \beta +\gamma \right)\mp \sqrt{\lambda _1^2 \left(4 \alpha _1^2+x_f^4 \left(3 \alpha _1-2 \beta +\gamma \right){}^2-4 \alpha _1 x_f^2 \left(3 \alpha _1+2 (\beta +\gamma )\right)\right)}}{\sqrt{6} \alpha _1 \lambda _1^2},x_2\to 0,
    \\
    y_2\to 0,u\to 0,
    \\
    z_1\to 0,z_2\to \frac{\pm \sqrt{\lambda _1^2 \left(4 \alpha _1^2+x_f^4 \left(3 \alpha _1-2 \beta +\gamma \right){}^2-4 \alpha _1 x_f^2 \left(3 \alpha _1+2 (\beta +\gamma )\right)\right)}+\lambda _1 \left(x_f^2 \left(3 \alpha _1-2 \beta +\gamma \right)-2 \left(\alpha _1+\gamma \right)\right)}{2 \lambda _1 \left(2 \alpha _1+\gamma \right)} \Bigg],
\end{multline}
with the effective equation of state:
\begin{equation}
    w_{eff}=\frac{\lambda _1 \left(x_f^2 \left(3 \alpha _1-2 \beta +\gamma \right)-\alpha _1\right) \mp\sqrt{\lambda _1^2 \left(4 \alpha _1^2+x_f^4 \left(3 \alpha _1-2 \beta +\gamma \right){}^2-4 \alpha _1 x_f^2 \left(3 \alpha _1+2 (\beta +\gamma )\right)\right)}}{3 \alpha _1 \lambda _1},
\end{equation}
describes a critical line where the kinetic energy of the axion field affects the dynamics, being influenced by different coupling parameters. Also, the dynamics is driven by the scalar curvature coupling of the saxion field. This critical line is also non-hyperbolic, with one zero eigenvalue. The complexity of the eigenvalues is high and the results cannot be written in the manuscript. As in the previous cases, by fine-tuning one can find various specific intervals when a saddle critical behavior is reached. For example, if we set $(x_f=1, \alpha_1=1, \alpha_2=1, \lambda_1=1, \gamma=1, \beta=-1$, we obtain a saddle behavior with the following eigenvalues: $P_4^{E}=[0, -1.64, -0.82 \pm 4.06i, 2.71, 2.71, -0.29]$. In Figs.\ref{fig:p4plusweff},\ref{fig:p4plusweffzero},\ref{fig:p4plusweffminusunu}, we display the effective equation of state for the $P_4^{+}$ solution, exhibiting an accelerated expansion scenario, a matter dominated epoch, and a de-Sitter evolution.

 \par
\begin{figure}[t]
  \includegraphics[width=8cm]{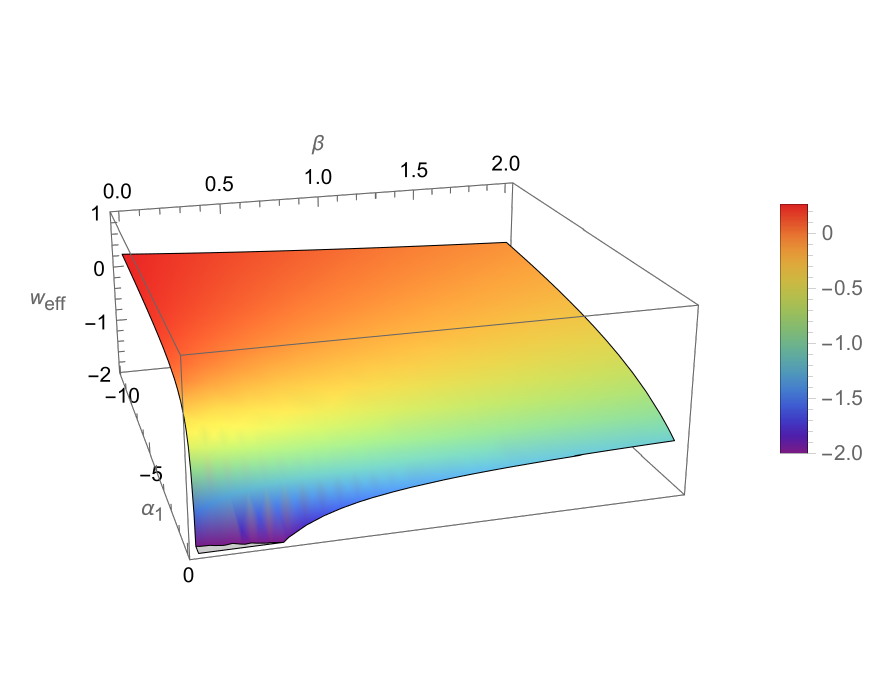} 
\caption{The effective equation of state for the $P_5^{-}$ solution ($\gamma=1$).}
\label{fig:p5minusweff}       
\end{figure}

 
 For the $P_5^{\pm}$ solution the location in the phase space structure is defined by the following coordinates:

 \begin{equation}
     P_5^{\pm}=\Bigg[ x_1\to \pm\frac{\sqrt{\frac{2}{3}} \left(2 \alpha _1+\gamma \right)}{\lambda _1 \left(2 \beta -\alpha _1\right)},x_f\to \pm\frac{\sqrt{\alpha _1}}{\sqrt{\alpha _1-2 \beta }},x_2\to 0,y_2\to 0,u\to 0,z_1\to 0,z_2\to \frac{\alpha _1-\gamma }{2 \alpha _1+\gamma }\Bigg],
 \end{equation}

 with the effective equation of state:

 \begin{equation}
     w_{eff}=\frac{\alpha _1+6 \beta +2 \gamma }{3 \alpha _1-6 \beta }.
 \end{equation}
 The value of the coupling constants affect the physical characteristics. We have plotted in Fig.\ref{fig:p5minusweff} the effective equation of state, spanning the corresponding interval relevant from a cosmological point of view. The eigenvalues for the $P_5^{-}$ solution are:

 \begin{equation}
     P_5^{-,E}=\Big[0,-\frac{2 \alpha _2 \left(2 \alpha _1+\gamma \right)}{2 \beta -\alpha _1},-\frac{2 \left(2 \alpha _1+\gamma \right)}{2 \beta -\alpha _1},-\frac{\alpha _1+6 \beta +2 \gamma }{2 \beta -\alpha _1}, E_5, E_6, E_7 \Big].
 \end{equation}
 As in the previous cases the eigenvalues have complicated expressions and are not displayed in the manuscript. We can fine tune the model's parameters in order to obtain a specific dynamical behavior. For example if we set $(\lambda_1=1, \gamma=1, \beta=1, \alpha_1=1)$ we obtain the following eigenvalues,

 \begin{equation}
     P_5^{-,E}=\Big[0.,-6. \alpha _2,-6.,-9.,-6.87298,-6.,0.872983 \Big],
 \end{equation}
corresponding to a saddle solution.

\par 
Next, the $P_6$ solution is driven by the kinetic energy of the axion field, and the specific function which couples the kinetic energy of the axion with the value of the saxion field. Hence,  we have the following location in the phase space structure, 

\begin{equation}
    P_6^{\pm}=\Bigg[x_1\to \frac{\sqrt{6} \gamma }{\lambda _1 (\gamma -2 \beta )},x_f\to \pm \frac{\sqrt{\gamma }}{\sqrt{\gamma -2 \beta }},x_2\to 0,y_2\to 0,u\to 0,z_1\to 0,z_2\to 0 \Bigg].
\end{equation}

The total equation of state has the following dependence, 

\begin{equation}
w_{eff}=\frac{2 \beta +\gamma }{\gamma -2 \beta },
\end{equation}
and can mimic different specific epochs by fine-tuning the value of the coupling parameters. The have obtained the following conditions for the acceleration $(w_{eff}<-\frac{1}{3})$, 

\begin{equation}
    (\beta <0\land 2 \beta <\gamma <-\beta )\lor (\beta >0\land -\beta <\gamma <2 \beta ).
\end{equation}

The corresponding eigenvalues for this solution,

\begin{equation}
    P_6^{+,E}=\Big[0,\frac{6 \text{$\alpha_2 $} \gamma }{\text{$\lambda_1 $} (\gamma -2 \beta )},\frac{6 \gamma }{\gamma -2 \beta },\frac{3 (2 \beta +\gamma )}{\gamma -2 \beta }, ... \Big]
\end{equation}
have an increased complexity. In the case when we set $(\beta=1, \gamma=1, \lambda_1=1)$ we obtain a saddle behavior characterized by the following values:

\begin{equation}
     P_6^{+,E}=\Big[0.,-6. \alpha _2,-6.,-9.,0.872983\, +0. i,-6.+0. i,-6.87298+0. i \Big].
\end{equation}

\begin{figure}[t]
  \includegraphics[width=8cm]{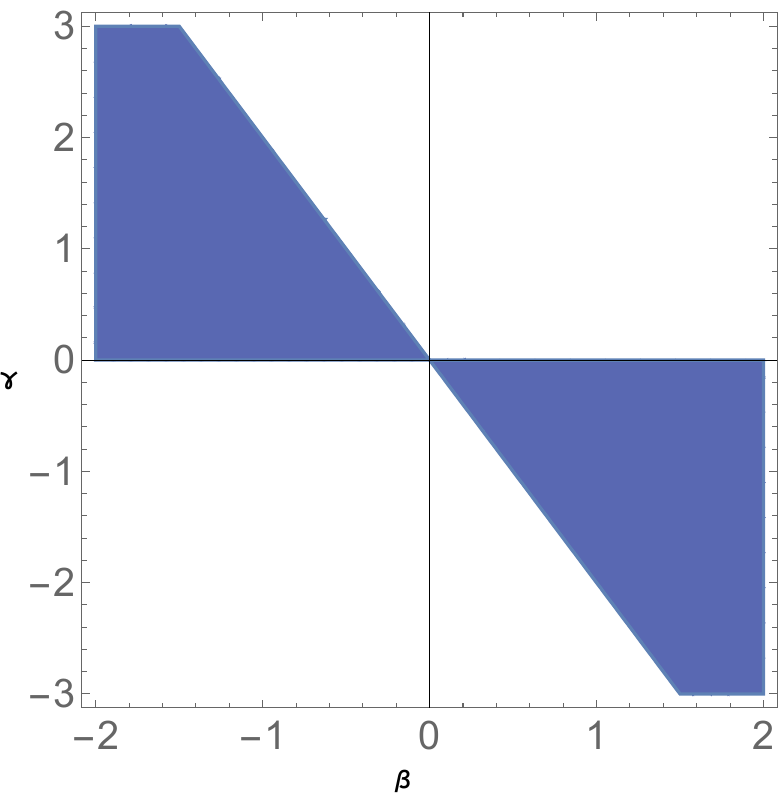} 
\caption{Saddle region for the $P_6^{+}$ solution ($\lambda_1=1, \lambda_2=1, \alpha_1=1, \alpha_2=1$).}
\label{fig:p6plussaddle}       
\end{figure}

Furthermore, for the $P_6^{+}$ solution we have depicted in Fig.\ref{fig:p6plussaddle} a non-exhaustive region where the dynamics corresponds to a saddle dynamical behavior.

\par 
The next solution $P_7^{\pm}$ is affected by the kinetic energy of the saxion field, together with the kinetic energy of the axion field, modulated by the specific coupling function. The location in the phase space is represented by, 

\begin{equation}
    P_7^{\pm}=\Bigg[x_1\to 0,x_f\to \pm \frac{\sqrt{-2 \beta  \gamma +\gamma ^2-6}}{\sqrt{(\gamma -2 \beta )^2}},x_2\to \frac{\sqrt{6}}{\gamma -2 \beta },y_2\to 0,u\to 0,z_1\to 0,z_2\to 0 \Bigg],
\end{equation}

with the corresponding equation of state,

\begin{equation}
    w_{eff}=-\frac{2 \beta +\gamma }{2 \beta -\gamma }.
\end{equation}

For the $P_7{+}$ solution we have obtained the following eigenvalues, 

\begin{multline}
     P_7^{+,E}=\Big[0,\frac{12 \beta }{\gamma -2 \beta }+3,\frac{6 \alpha _1}{\gamma -2 \beta },-\frac{6 \beta }{\gamma -2 \beta }, -\frac{6 \left(\gamma -\lambda _2\right)}{2 \beta -\gamma }, \\
     \frac{\pm\sqrt{3} \sqrt{-2 \beta  \gamma +\gamma ^2-6} \sqrt{\Xi}-3 \beta  \left(-2 \beta  \gamma +\gamma ^2-6\right)^2 \sqrt{(\gamma -2 \beta )^2} (2 \beta -\gamma )^5}{(\gamma -2 \beta )^6 \sqrt{(\gamma -2 \beta )^2} \left(-2 \beta  \gamma +\gamma ^2-6\right)^2}\Big],
\end{multline}

with 
\begin{equation}
    \Xi=-\beta  (\gamma -2 \beta )^{12} \left(2 \beta  \gamma -\gamma ^2+6\right)^3 \left(8 \beta ^2 \gamma +\beta  \left(27-8 \gamma ^2\right)+2 \gamma  \left(\gamma ^2-6\right)\right).
\end{equation}

\begin{figure}[t]
  \includegraphics[width=8cm]{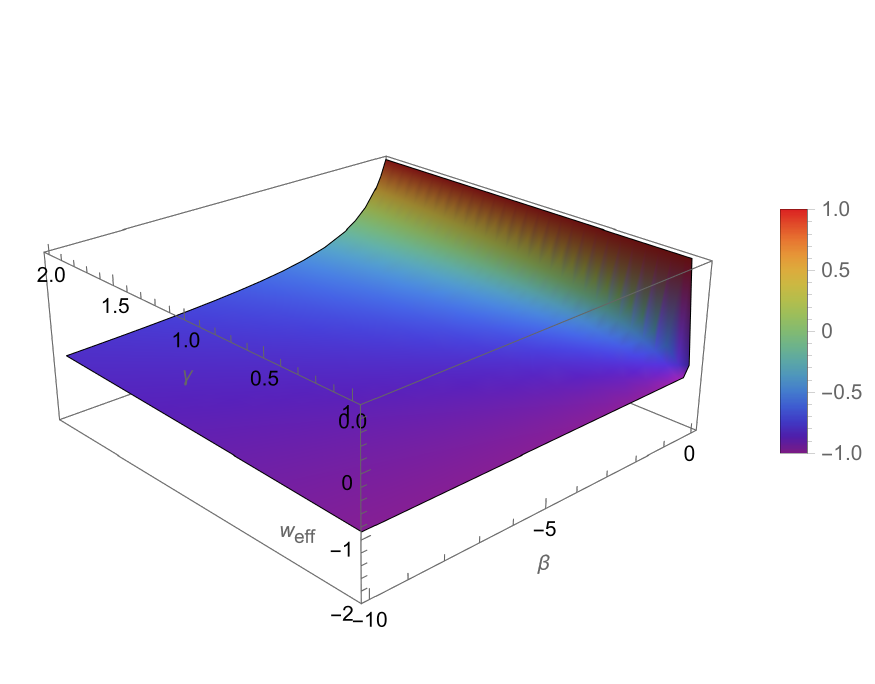} 
\caption{The variation of the total equation of state for the $P_7$ solution.}
\label{fig:p7weff}       
\end{figure}

\begin{figure}[t]
  \includegraphics[width=8cm]{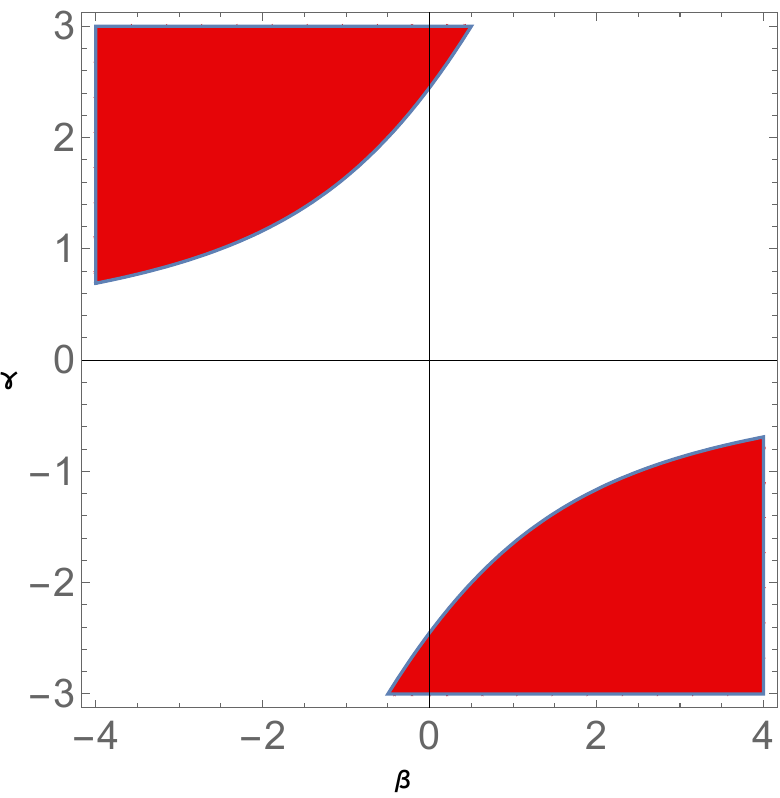} 
\caption{Saddle region for the $P_7^{+}$ solution ($ \lambda_2=1, \alpha_1=1$).}
\label{fig:p7plussaddle}       
\end{figure}

\begin{figure}[t]
  \includegraphics[width=8cm]{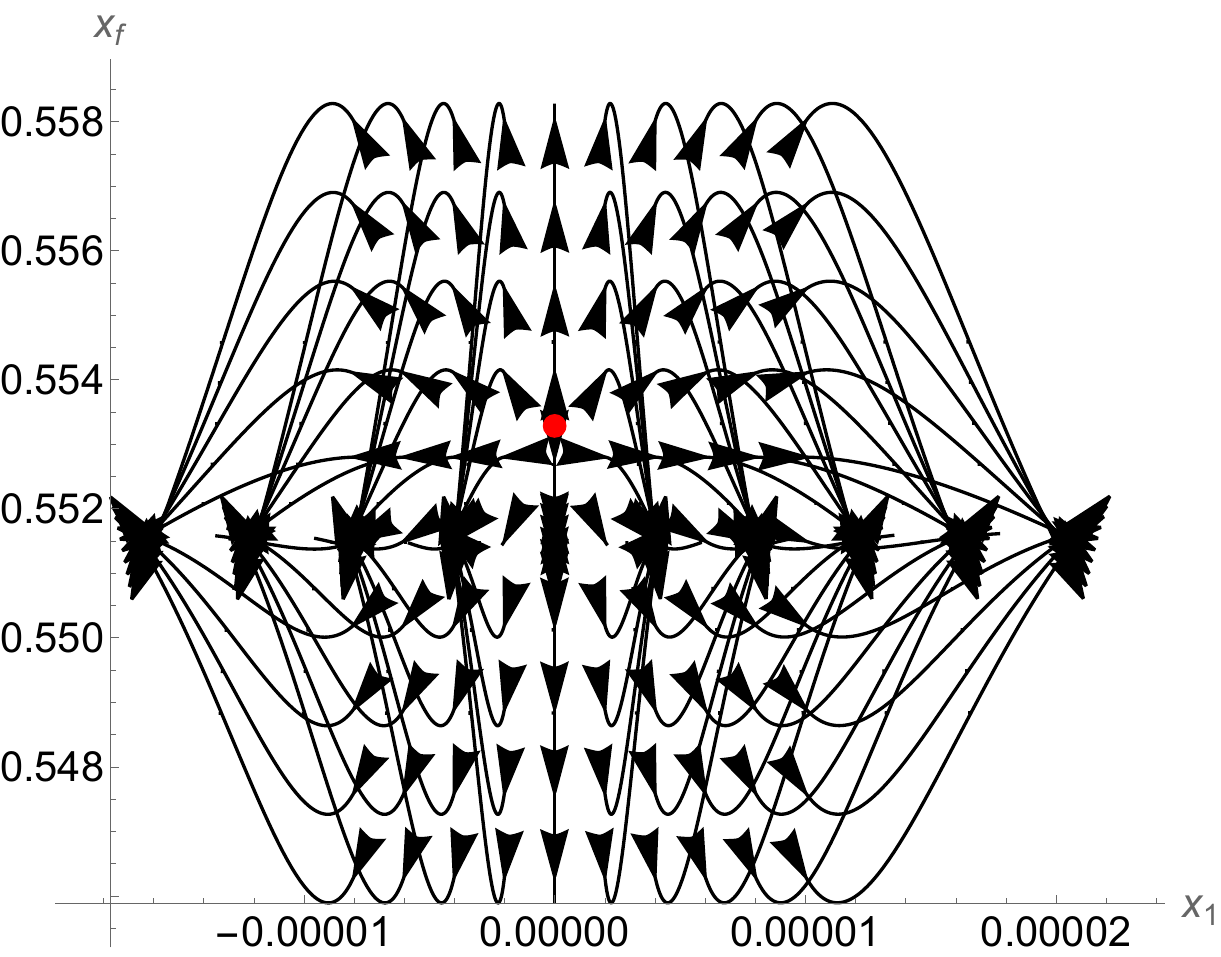} 
\caption{Numerical evolution near the  $P_7^{+}$ solution. ($ \lambda=1, \alpha=1, \gamma=-3, \beta=2$).}
\label{fig:p7plusnumerics}       
\end{figure}

By proper fine tuning we can obtain the relevant values for different cosmological epochs, as seen in Fig.~\ref{fig:p7weff}. Hence, this critical point can mimic specific cosmological eras, like matter, radiation, or de Sitter. A viable solution where the behavior corresponds to a saddle era is described in Fig.~\ref{fig:p7plussaddle}. Moreover, in Fig.~\ref{fig:p7plusnumerics} we show that in certain situations this solution can act as a repeller, confirming the viability between the analytical solutions and the numerical aspects.

\par 
For the $P_8{\pm}$ solution we have obtained the following coordinates, 

\begin{equation}
    P_8^{\pm}=\Bigg[ x_1\to 0,x_f\to \pm \frac{\sqrt{-2 \beta  \lambda _2+\lambda _2^2-6}}{\sqrt{\left(\lambda _2-2 \beta \right){}^2}},x_2\to -\frac{\sqrt{6}}{2 \beta -\lambda _2},y_2\to \frac{2 \beta }{2 \beta -\lambda _2},u\to 0,z_1\to 0,z_2\to 0\Bigg],
\end{equation}

with 

\begin{equation}
    w_{eff}=-\frac{2 \beta +\lambda _2}{2 \beta -\lambda _2},
\end{equation}

while for the eigenvalues we have (for the $P_8^{+}$), 

\begin{equation}
     P_8^{+,E}=\Big[ 0,-\frac{3 \left(2 \beta +\lambda _2\right)}{2 \beta -\lambda _2},-\frac{6 \alpha _1}{2 \beta -\lambda _2},\frac{6 \beta }{2 \beta -\lambda _2}, ... \Big].
\end{equation}

As in the previous cases, the last eigenvalues are too complex to be written here. The effective equation of state is similar to the one specific for the $P_2$ solution, but the location in the phase space structure is quite different. In the case when we fix the parameters with $(\beta = 2,\gamma = -3,\lambda_2 = -1)$ we obtain a saddle evolution characterized by the following eigenvalues: 

\begin{equation}
     P_8^{+,E}=\Big[\{0.,-1.8,-1.2 \text{$\alpha $1},2.4,0.759592\, +0. i,-2.4+0. i,-3.15959+0. i\}\Big].
\end{equation}

\par 
The remaining cosmological solutions of the autonomous system of equations are not further discussed since these critical points denoted as $P_{9,10,11,12}$ are specific to a stiff evolution ($w_{eff}=+1$) and are not so relevant for the modern cosmology. However, for completeness we have added the specific eigenvalues for these cases:

\begin{equation}
     P_9^{E}=\Big[0,0,-\frac{i \sqrt{6} \sqrt{\beta } \sqrt{\alpha _1-2 \beta } x_f}{\sqrt{6 \beta ^2 x_f^2+1}},\frac{i \sqrt{6} \sqrt{\beta } \sqrt{\alpha _1-2 \beta } x_f}{\sqrt{6 \beta ^2 x_f^2+1}},3,6,6\Big],
\end{equation}

\begin{equation}
     P_{10}^{+,E}=\Big[0,-\frac{i \sqrt{6} \sqrt{\alpha _1} \sqrt{\beta } \sqrt{\alpha _1-2 \beta }}{\sqrt{\alpha _1 \left(6 \beta ^2+1\right)-2 \beta }},\frac{i \sqrt{6} \sqrt{\alpha _1} \sqrt{\beta } \sqrt{\alpha _1-2 \beta }}{\sqrt{\alpha _1 \left(6 \beta ^2+1\right)-2 \beta }},3,6,\sqrt{6} \alpha _2 x_1,6-\sqrt{6} \lambda _1 x_1 \Big],
\end{equation}

\begin{equation}
     P_{11}^{+,E}=\Big[ 0,0,-\frac{i \sqrt{6} \sqrt{\alpha _1} \sqrt{\beta } \sqrt{\alpha _1-2 \beta }}{\sqrt{\alpha _1 \left(6 \beta ^2+1\right)-2 \beta }},\frac{i \sqrt{6} \sqrt{\alpha _1} \sqrt{\beta } \sqrt{\alpha _1-2 \beta }}{\sqrt{\alpha _1 \left(6 \beta ^2+1\right)-2 \beta }},3,6,6\Big],
\end{equation}

\begin{equation}
     P_{12}^{+,E}=\Big[ 0,\frac{2 \alpha _2 \left(2 \alpha _1+\gamma \right)}{\lambda _1 \left(\alpha _1-2 \beta \right)},-\frac{i \sqrt{6} \sqrt{\alpha _1} \sqrt{\beta } \sqrt{\alpha _1-2 \beta }}{\sqrt{6 \alpha _1 \beta ^2+\alpha _1-2 \beta }},\frac{i \sqrt{6} \sqrt{\alpha _1} \sqrt{\beta } \sqrt{\alpha _1-2 \beta }}{\sqrt{6 \alpha _1 \beta ^2+\alpha _1-2 \beta }},3,6,\frac{2 \left(\alpha _1-6 \beta -\gamma \right)}{\alpha _1-2 \beta }\Big].
\end{equation}

\par
\begin{figure}[t]
  \includegraphics[width=8cm]{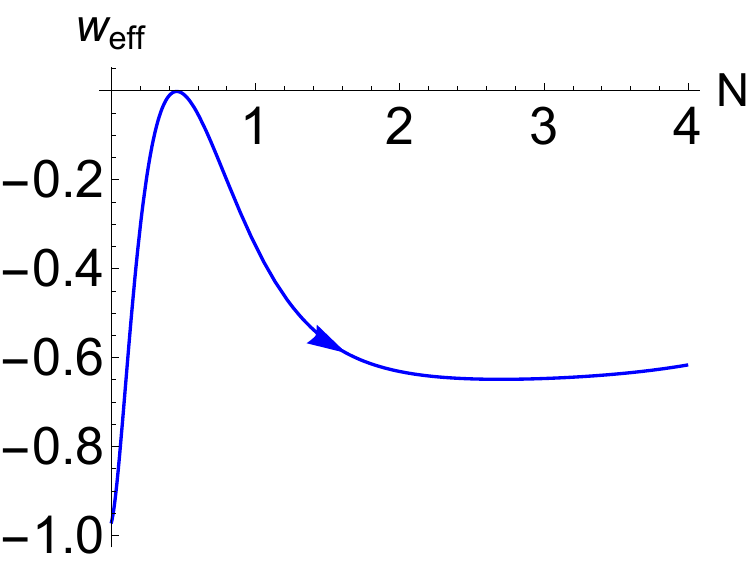} 
\caption{Numerical evolution of the present cosmological model. }
\label{fig:numerics}       
\end{figure}

 
Lastly, in Fig.~\ref{fig:numerics} we have displayed a possible evolution in the present cosmological system, characterized by a quintessence behavior. By fine-tuning, we can obtain various trajectories in the phase space structure, confirming the richness of the current cosmological scenario. This includes also the mimic of the matter domination epoch, since our model has a scaling solution which can alleviate the cosmic coincidence problem.

\begin{sidewaystable}[ht]
\centering
\begin{center}
\scalebox{0.65}{
\begin{tabular}{||c c c c c c c c||} 
 \hline
 P & $x_1$ & $x_2$ & $x_f$ & $y_2$ & $u$ & $z_1$ & $z_2$ \\ [0.5ex] 
 \hline
 $P_1$ & 0 & $\frac{\sqrt{6}}{\gamma -2 \beta }$ & $x_f$ & $0$ & $0$ & $\frac{-2 \beta  \gamma +\gamma ^2-4 \beta ^2 x_f^2+4 \beta  \gamma  x_f^2+\gamma ^2 \left(-x_f^2\right)-6}{\gamma  (2 \beta -\gamma )}$ & $0$ \\ 
 \hline
  \hline
 $P_2$ & $0$ & $-\frac{\sqrt{6}}{2 \beta -\lambda _2}$ & $x_f$ & $-\frac{2 \left(6 \beta +4 \beta ^3 x_f^2-4 \beta ^2 \lambda _2 x_f^2+\beta  \lambda _2^2 x_f^2\right)}{\lambda _2 \left(2 \beta -\lambda _2\right){}^2}$ & $0$ & $\frac{-2 \beta  \lambda _2-4 \beta ^2 x_f^2+4 \beta  \lambda _2 x_f^2-\lambda _2^2 x_f^2+\lambda _2^2-6}{\lambda _2 \left(2 \beta -\lambda _2\right)}$ & $0$ \\ 
 \hline
  \hline
 $P_3$ & $\frac{\sqrt{\frac{3}{2}}}{\lambda _1}$ & $0$ & $x_f$ & $0$ & $\frac{\alpha _1-2 \alpha _1 x_f^2-4 \beta  x_f^2-2 \gamma  x_f^2}{\alpha _1}$ & $0$ & $-\frac{2 (2 \beta +\gamma ) x_f^2}{\alpha _1} $ \\ 
 \hline
  \hline
 $P_4^{\pm}$ & $\frac{2 \alpha _1 \lambda _1+\lambda _1 x_f^2 \left(3 \alpha _1-2 \beta +\gamma \right)\mp \sqrt{\lambda _1^2 \left(4 \alpha _1^2+x_f^4 \left(3 \alpha _1-2 \beta +\gamma \right){}^2-4 \alpha _1 x_f^2 \left(3 \alpha _1+2 (\beta +\gamma )\right)\right)}}{\sqrt{6} \alpha _1 \lambda _1^2}$ & $0$ & $x_f$ & $0$ & $0$ & $0$ & $\frac{\pm \sqrt{\lambda _1^2 \left(4 \alpha _1^2+x_f^4 \left(3 \alpha _1-2 \beta +\gamma \right){}^2-4 \alpha _1 x_f^2 \left(3 \alpha _1+2 (\beta +\gamma )\right)\right)}+\lambda _1 \left(x_f^2 \left(3 \alpha _1-2 \beta +\gamma \right)-2 \left(\alpha _1+\gamma \right)\right)}{2 \lambda _1 \left(2 \alpha _1+\gamma \right)} $ \\ 
 \hline
  \hline
 $P_5^{\pm}$ & $\pm\frac{\sqrt{\frac{2}{3}} \left(2 \alpha _1+\gamma \right)}{\lambda _1 \left(2 \beta -\alpha _1\right)}$ & $0$ & $\pm\frac{\sqrt{\alpha _1}}{\sqrt{\alpha _1-2 \beta }}$ & $0$ & $0$ & $0$ & $\frac{\alpha _1-\gamma }{2 \alpha _1+\gamma }$ \\ 
 \hline
  \hline
 $P_6^{\pm}$ & $\frac{\sqrt{6} \gamma }{\lambda _1 (\gamma -2 \beta )}$ & $0$ & $\pm \frac{\sqrt{\gamma }}{\sqrt{\gamma -2 \beta }}$ & $0$ & $0$ & $0$ & $0$ \\ 
 \hline
  \hline
 $P_7$ & $0$ & $\frac{\sqrt{6}}{\gamma -2 \beta }$ & $\frac{\sqrt{-2 \beta  \gamma +\gamma ^2-6}}{\sqrt{(\gamma -2 \beta )^2}}$ & $0$ & $0$ & $0$ & $0$ \\ 
 \hline
  \hline
 $P_8^{\pm}$ & $0$ & $-\frac{\sqrt{6}}{2 \beta -\lambda _2}$ & $\pm \frac{\sqrt{-2 \beta  \lambda _2+\lambda _2^2-6}}{\sqrt{\left(\lambda _2-2 \beta \right){}^2}}$ & $\frac{2 \beta }{2 \beta -\lambda _2}$ & $0$ & $0$ & $0$ \\ 
 \hline
  \hline
 $P_9$ & 0 & 0 & $x_f$ & 0 & 0 & $\frac{-\alpha _1+\alpha _1 x_f^2-2 \beta  x_f^2}{\alpha _1}$ & $\frac{2 \beta  x_f^2}{\alpha _1}$ \\ 
 \hline
  \hline
 $P_{10}^{\pm}$ & $x_1$ & 0 & $\pm \frac{\sqrt{\alpha_1}}{\sqrt{\alpha_1-2 \beta}}$ & 0 & 0 & 0 & $\frac{2 \beta }{\alpha_1-2 \beta }$ \\ 
 \hline
  \hline
 $P_{11}^{\pm}$ & 0 & 0 & $\pm \frac{\sqrt{\alpha_1}}{\sqrt{\alpha_1-2 \beta}}$ & 0 & 0 & 0 & $\frac{2 \beta}{\alpha_1-2 \beta}$ \\ 
 \hline
  \hline
 $P_{12}^{\pm}$ & $\frac{2 (2 \alpha_1+\gamma)}{\sqrt{6}\alpha_1 \lambda_1 - 2 \sqrt{6} \beta \lambda_1}$ & 0 & $\pm \frac{\sqrt{\alpha_1}}{\sqrt{\alpha_1-2 \beta}}$ & 0 & 0 & 0 & $\frac{2 \beta}{\alpha_1-2 \beta}$ \\ 
 \hline
\end{tabular}}
\end{center}
\caption{The critical points of the autonomous system of equations.}
\label{tab:tab1}
\end{sidewaystable}

\section{Conclusions}
\label{conclusions}
\par 
In this paper we have extended a recent axion-dilaton cosmological system by considering viable influences from the space-time geometry, through a non-minimal coupling with gravity. The cosmological system is endowed with a possible interplay between the scalar fields and the space-time geometry, by taking into account the scalar curvature. Hence, the two scalar fields which are describing the dark sector are non-minimally coupled in an independent manner with the Ricci scalar.
\par 
In our approach, the saxion-like field is a scalar field having a trivial kinetic energy, while the axion-like component is endowed with a nontrivial kinetic energy modulated by the value of the saxion-like field through a specific function. The particular extension considered in the present manuscript is based on a non-minimal coupling of the two scalar fields with gravity through the most simple invariant component, the scalar curvature. After proposing the action for our cosmological model, we have obtained the modified Friedmann relations, which satisfy the usual continuity equation, since the extension is not affected by different invariants based on the energy-momentum tensor. The physical implications for the current cosmological model are analyzed using linear stability theory, an important analytical tool that associates a dynamical system to the system of equations. For this system we have obtained the corresponding critical points, which can describe various epochs in the evolution of our Universe, like radiation, matter domination, de Sitter, and so on. 
\par 
In our case the analysis revealed the existence of various critical points which can be associated with the previous mentioned epochs. For each critical point, we have obtained the associated eigenvalues which describe the dynamical properties of the cosmological system. In this case, we have obtained specific relations associated with various physical effects, obtaining possible analytical constraints based on dynamical analysis. For this model, we have obtained also a specific scaling solution, a critical point associated to the existence of a possible matter-dominated epoch, alleviating the cosmic coincidence problem for cosmological systems. 
\par
The present cosmological model can be extended in various theoretical directions. A possible direction is related to the geometrical invariants, going beyond the scalar curvature, and considering different terms like the Gauss-Bonnet topological component, or different new invariants based on cubic contractions of the Riemann tensor. Furthermore, an important analysis would be associated to the cosmological observations, by confronting the theoretical values of different variables near the de-Sitter critical points with observed data acquired through various cosmological observations. Finally, the present cosmological model can be embedded in different modified gravity theories, based on scalar curvature, the Gauss-Bonnet term, and so on. All of these directions are open and left for future projects.

\bibliography{sorsamp}

\end{document}